\documentclass{kluwer}

\newcommand{\Emax}{E_{\rm max}}
\newcommand{\Vsk}{V_{\rm sk}}
\newcommand{\Rsk}{R_{\rm sk}}
\newcommand{\EnSN}{E_{\rm SN}}
\newcommand{\Mej}{M_{\rm ej}}
\newcommand{\tSNR}{t_{\rm snr}}
\newcommand{\tacc}{t_{\rm acc}}

\begin{document}


\begin{article}

\begin{opening}

\title{Tests of Galactic Cosmic Ray Source Models}
\subtitle{Report of working group 4}



\author{L. O'C. \surname{Drury}}
\author{D. C. \surname{Ellison}}
\author{F. A. \surname{Aharonian}}
\author{E. \surname{Berezhko}}
\author{A. \surname{Bykov}}
\author{A. \surname{Decourchelle}}
\author{R. \surname{Diehl}}
\author{G. \surname{Meynet}}
\author{E. \surname{Parizot}}
\author{J. \surname{Raymond}}
\author{S. \surname{Reynolds}}
\author{S. \surname{Spangler}}

\end{opening}

\section{Introduction}

The problem of understanding the origin of the Galactic Cosmic Rays
(GCRs) is an old and recalcitrant one. It is actually several distinct
problems. First, there is the question of the origin of the energy.
What powers the accelerator and how does it work?  Second, there is
the question of the origin of the particles which are accelerated. Out
of what component of the Galaxy does the accelerator select particles
to turn into cosmic rays?  Third, there is the question of how much of
the observed cosmic ray spectrum is in fact of Galactic origin.  Over
what energy range does the accelerator work and what spectral form
does its output have? Finally, there is the question of how many
different types of accelerator are required. Can one basic process
explain all the data, or do we need to invoke multiple sources and
mechanisms? Of course a satisfactory physical model for the origin of
the GCRs should simultaneously answer all these questions, however, in
the context of looking for observational tests, it is sensible to
adopt a ``divide and conquer'' strategy and regard them as separate
questions.

The only theory of particle acceleration which at present is
sufficiently well developed and specific to allow quantitative model
calculations, and which appears capable of meeting many of the
observational constraints on any cosmic ray acceleration theory, is
diffusive acceleration applied to the strong shocks associated with
supernova remnants.  Thus this report concentrates, {\it faute de
mieux}, on tests of this hypothesis, described in more detail in the
next section.

\section{SNR shocks as sources of the GCR energy}

The fact that the power required to maintain the GCR population is
estimated as a few to several percent of the mechanical energy input
to the Galaxy from SNe explosions, together with a distinct lack of
other plausible energy sources (with the possible exception of
gamma-ray bursts, which also meet energy requirement), is a strong
hint that the ultimate power source for the GCR accelerator is to be
found in SNe.  However if the GCR were accelerated in the explosion
itself, the adiabatic losses experienced by the GCR particles in
pushing aside the ambient interstellar medium (ISM) would raise the
energy requirements to an impossible level.  Thus the acceleration
site must be located in the subsequent supernova remnant (SNR) and in
diffusive shock acceleration we have a convincing mechanism for doing
this.

\subsection{Predictions of nonlinear nonthermal shock models of SNRs}

There have been substantial developments in our understanding of
diffusive shock acceleration, especially as applied to SNR shocks, in
the last several years (e.g., Berezhko \& Voelk 2000; Berezhko \&
Ellison 1999; Ellison et al. 1997; Meyer et al. 1997). The key advance
has been improved understanding of the nonlinear reaction effects of
the accelerated particles on the shock structure, an essential aspect
if the process is to operate with high efficiency.  One of the most
promising aspects of this work is that, despite the uncertainties and
the ad-hoc assumptions that still have to be made, there appears to be
good agreement between the different approaches.  Specific predictions
of all nonlinear nonthermal shock models ranging from simple fluid
models through various Monte Carlo and kinetic models to asymptotic
analytic theories are the following:

\begin{itemize}

\item An extended precursor region on the upstream side of the shock
in which the material flowing into the shock is decelerated,
compressed and heated and where the magnetic field is strongly
disturbed.

\item A subshock, essentially a conventional shock, marking the transition
from the upstream precursor region to the downstream region;
relative to shock models without particle acceleration
the overall compression ratio is significantly enhanced, but the
subshock ratio is reduced.

\item Lower postshock temperatures downstream (relative to shock
models without particle acceleration), but preheating of the bulk
plasma on the upstream side of the subshock both through adiabatic
compression and dissipation of the intense magnetic turbulence.

\item Accelerated particle momentum distribution functions which are
close to power-law, but slightly concave and ``hard'',
over an extended region between the thermal peak and the upper
cut-off.

\item Thermal and non-thermal ion populations which join smoothly
through non-Maxwellian tails on the quasi-thermal shocked distribution
and an energy content in the accelerated particle population which is
a significant part of the shock energy budget.

\end{itemize}
The length and time scales associated with the precursor structure are
determined by the diffusion of the accelerated particles in the shock
neighbourhood. It is important to note that, because of the strongly
perturbed magnetic field, the diffusion coefficients are very much
smaller than in the general interstellar medium. The usual assumption
which is made is that the diffusion obeys Bohm scaling with a mean
free path which is of order the particle gyro-radius. If particles are
efficiently accelerated up to the maximum energy allowed by the
geometry and age of the shock, the precursor length-scale for the
highest energy particles will typically be of order one tenth the
shock radius and proportionally less at lower energies.

\subsection{Observational tests of SNR source models}

\subsubsection{Radio diagnostics}

Radio observations of electron synchrotron emission, because of the
excellent sensitivity and angular resolution of modern radio
telescopes, are powerful probes of the distributions of relativistic
electrons and magnetic fields in and around SNRs.  Unfortunately the
magnetic fields are usually only poorly known and this considerably
complicates the interpretation of the radio data in isolation.
Another problem is that, since the characteristic emission frequency
scales as the electron energy squared, a very wide spectral range of
synchrotron emission must be observed to learn about any appreciable
energy range of the electron spectrum. And of course the observations
tell us nothing directly about accelerated {\em ion}
populations. However, for conventional magnetic fields of a few $\mu$G
and observing frequencies in the GHz range the emission is dominated
by electrons of a few GeV energy, so the radio studies typically
sample electrons of comparable energies and rigidities to the mildly
relativistic protons which dominate the energy density of the Galactic
cosmic rays and there is not reason to suppose that at these energies
the transport and acceleration of electrons is very different to that
of protons.

The simple test-particle theory of shock acceleration gives a fixed
relation between the shock compression ratio $r$ and the electron
differential energy spectral index, $s=(r+2)/(r-1)$, which in turn is
related to the synchrotron spectral index, $\alpha \equiv (s-1)/2$. If
the equation of state is close to that of an ideal gas with a ratio of
specific heats $\gamma = 5/3$, then all strong shocks have $r \approx
4$ and thus $s = 2$, implying synchrotron spectral indices $\alpha =
0.5$.  At radio frequencies, values below $0.5$, commonly observed
among Galactic shell remnants, then either require lossy shocks with
compression ratios above 4 (either radiative shocks or efficient
cosmic-ray-accelerating shocks), or confusion with flat-spectrum
thermal radio emission.  Values above 0.5 in the test-particle picture
require weak shocks, unacceptably so for young remnants such as Tycho
or Kepler (data in Green, 1998), or curved spectra (hardening to higher
energies) as predicted by nonlinear acceleration models \cite{Ellison91}.
The curvature not only gives direct evidence for a modified shock and
electron diffusion coefficient increasing with energy, but can in
principle be used to find the mean magnetic-field strength \cite{Reynolds92},
though in practice the data are not of high enough quality to enable
this.

Two areas of concern exist for this predicted curvature.  First, in
Cas A, the integrated radio spectral index is constant at about
$-0.78$ from about 10 MHz to 100 GHz -- a factor of 100 in electron
energies, over which the spectrum is predicted to have measurable
curvature (e.g., Ellison et al.  2000).  Second, in radio supernovae,
spectral indices of up to 1.0 are observed, which do not change with
time as would be expected if increasing shock modification by
energetic particles is occurring (Weiler et al., 1990; Gaensler et
al., 1997; Montes et al., 1998, 2000).

In diffusive shock acceleration, electrons diffuse a significant
distance ahead of the shock in the process of gaining energy -- far
enough ahead to produce a potentially observable synchrotron precursor
in the radio.  Achterberg, Blandford, \& Reynolds (1994) used this
effect, in conjunction with observations of several sharp-rimmed SNRs,
to put a lower limit on the upstream electron diffusion length.  They
conclude that MHD turbulence upstream of shocks in four young SNRs
must have amplitudes larger than those responsible for average
galactic cosmic-ray diffusion (near 1 GeV) by factors of at least 60.
Unfortunate magnetic-field geometry in all four cases could render
larger precursors invisible, but alignment of the external magnetic
field to less than $\sim 30^{\circ}$ of the line of sight would be
required in all cases.  This effect can be searched for in all radio
observations of sharp-rimmed remnants; applications in larger,
presumably older remnants (e.g., CTA 1; Pineault et al.~1997;
Aschenbach \& Leahy 1999) require intermediate levels of enhanced
turbulence.  In no case has a structure been seen in a radio SNR image
which can be unambiguously identified as pre-shock synchrotron
emission, though dimensionless amplitudes $\delta B/B ~ \sim 0.1$ are
adequate to render radio halos too thin to resolve.  For remnants
whose X-ray emission is dominantly synchrotron, halos must be
observably large, but may be too faint to detect.

\subsubsection{Optical and Ultraviolet diagnostics}

Radiative shocks efficiently convert thermal energy to radiation in a
cooling zone far downstream from the shock, and there most signatures
of physical processes in the shock have been erased.  More interesting
are non-radiative shocks in which the gas does not cool
appreciably after being shocked.  In this case optical and UV lines
are emitted from the narrow layer where the gas is ionized just behind
the shock.  These shocks are faint, but have been detected in about a
dozen SNRs; UV imaging should show the positions of similar shocks in
other remnants.  The faint emission from non-radiative shocks yields
important limits on electron-ion and ion-ion temperature equilibration
and can also be used to investigate the precursor predicted by
diffusive shock acceleration models.

The most complete diagnostics are available for shocks in partially
neutral gas (Chevalier \& Raymond 1978).  A neutral hydrogen atom
feels neither plasma turbulence nor electromagnetic fields as it
passes through a collisionless shock.  It will be quickly ionized in
the hot post-shock gas, but it may be excited and produce a photon
first.  On average, it will produce $q_{ex} / q_{i}$ photons (the
excitation rate over the ionization rate) before being destroyed, or
about 0.2 H$\alpha$ photons per $\rm H^0$ atom.  Because the neutrals
do not feel the shock itself, the H$\alpha$ profile reveals the
pre-shock velocity distribution of the H atoms.  However, there is a
substantial probability that the atom will undergo a charge transfer
reaction before being ionized.  This produces a population of neutrals
having a velocity distribution similar to that of the post-shock ions,
and they produce a corresponding broad component to the H$\alpha$
profile.

Thus the line widths of the broad and narrow components of H$\alpha$
measure the post-shock and pre-shock proton kinetic temperatures quite
directly.  In a few cases, UV lines of He II, C IV, N V and O VI are
also detectable.  These ions are affected by the shock, and the line
widths directly measure the kinetic temperatures of those species.  In
the two cases measured so far, the UV lines imply roughly
mass-proportional temperatures (Raymond et al. 1995; Raymond et
al. 2000).

There are two ways to find $\rm T_e$ in these shocks.  The intensity
ratio of the broad and narrow components of H$\alpha$ depends on the
ratio of the charge transfer rate to the ionization rate, $\rm q_{ct}
/ q_i$, and the latter depends of $\rm T_e$.  Ghavamian (1999) finds
that $\rm T_e / T_p$ just behind the shock varies from more than 80\%
in the 350 km/s shock in the Cygnus Loop to 40-50\% in the 620 km/s
shock in RCW 86 and less than 20\% in the 1800 km/s shock in Tycho.
Another determination of $\rm T_e / T_p$ used the UV lines in SN1006
(Laming et al. 1996).  Here, the C IV, N V and O VI excitation rates
are dominated by proton collisions, while the He II $\lambda$1640 line
is excited by electrons.  The observed line ratios imply $\rm T_e /
T_p < 0.2$.  In both the Cygnus Loop and SN1006, the parameters
derived from UV and optical lines agree well with analyses of the
X-ray spectra.  The tendency for a decreasing degree of electron-ion
equilibration with increasing shock speed matches results for shocks
in the solar wind (Schwartz et al. 1988).

Diffusive shock acceleration requires a precursor in which particles
scatter back and forth between the shock jump and MHD turbulence (e.g
Blandford and Eichler 1987).  Dissipation of the turbulence will heat
and accelerate the gas upstream of the shock jump in a precursor.  The
only detailed model of a precursor in partially neutral gas is that of
Boulares and Cox (1988).

The cosmic ray acceleration precursor should have four potentially
observable effects on UV and optical lines: 1) The narrow component of
H$\alpha$ will show the precursor temperature and turbulence rather
than ambient ISM values;\footnote{This assumes there is time for the
neutrals to feel the increasing temperature in the precursor.} 2)
Ionization in the precursor may reduce the hydrogen neutral fraction,
and cut down the flux of neutrals reaching the shock; 3) A shock
observed face-on should show a Doppler shifted narrow H$\alpha$ line;
and 4) Compression and heating in the precursor may produce faint
(narrow) H$\alpha$, [N II] and [S II].

Observationally, 1) the H$\alpha$ narrow components are 30-50 km/s
wide (Hester, Raymond and Blair 1994; Smith et al.  1994; Ghavamian et
al 2000), much too wide to be an ambient ISM temperature, 2) the
hydrogen neutral fraction required to match the observed H$\alpha$
brightness in the few cases analyzed limits the precursor thickness to
$\sim 10^{16-17}$ cm (approximately the upstream diffusion length of a
$10^{12}$ eV proton assuming $B\simeq 5\mu$G, a shock speed of 1000
km/s, and strong scattering), 3) no Doppler shift is seen in long slit
echelle image across the middle of an LMC Balmer-dominated SNR (Smith
et al.  1994), and, 4) faint N II] and [S II] detected at a Balmer
line filament in the NE Cygnus Loop by Fesen and Itoh (1985) may arise
in a precursor, while more extended emission (e.g. Bohigas, Sauvageot
and Decourchelle 1999; Szentgyorgyi et al 2000) is likely to be a
photoionization precursor.

Overall, the precursor needed in cosmic ray acceleration models
matches the observations except for the lack of Doppler shift.  An
alternative explanation is a precursor due to escape of the fast
component neutral hydrogen (produced by charge transfer with
post-shock protons) out the front of the shock.  This idea has not yet
been developed in detail.  The heating suggested by the observed
H$\alpha$ profiles is smaller than that in the Boulares and Cox (1988)
model, perhaps because Boulares and Cox assumed very efficient cosmic
ray acceleration.

\subsubsection{X-ray diagnostics}

Several SNRs are now known whose X-ray emission shows strong evidence
for the presence of nonthermal emission: SN 1006 (Koyama et al.~1995),
G347.3-0.5 (Koyama et al.~1997; Slane et al.~1999), RXJ~0852.0-0462
(Slane et al.~2001), Cas A (Allen et al.~1997; Favata et al.  1997;
The et al.~1996), RCW 86 (Borkowski et al.~2001).  In SN 1006,
G347.3-0.5 and RXJ~0852.0-0462, the X-ray spectrum is almost
featureless; Cas A shows many X-ray lines but a power-law continuum up
to 80 keV; and RCW 86 shows anomalously weak lines best explained as a
synchrotron continuum diluting a thermal spectrum.  Nonthermal
bremsstrahlung was suggested as a possible contributing process in Cas
A (Asvarov et al.~1990; Laming~2001).  However, in the former cases a
synchrotron explanation is preferred to nonthermal bremsstrahlung,
because of steepening or faint or nonexistent spectral lines.  This
synchrotron continuum can then provide powerful diagnostics of shock
acceleration.  However, nonthermal bremsstrahlung should become the
dominant source of photons above some energy, and future observations
should be able to discriminate.

\paragraph{Modifications of the thermal emission}

For nonlinear shock acceleration the overall shock compression is
increased.  Thus higher downstream densities are expected, which can
be derived in principle from the X-ray emission volume and ionization
timescale. However, the difficulty is then to distinguish between a
higher density ambient medium and a higher compression ratio. For
that, an independent determination of the upstream density is
required.  A possible way is to derive the upstream density from the
flux in the photoionized region surrounding some supernova remnants as
was done by Morse et al. (1996) for N132D using optical
observations. Other global information may give an indication of the
upstream density like the location in the galaxy (high latitude for SN
1006 and Kepler) or the study of the environs of the remnant
(e.g. Reynoso et al. 1999 for Tycho).

Efficient acceleration also lowers the downstream temperature, and the
post-shock electron temperature can be estimated from the X-ray
spectra.  With no acceleration effects the dimensionless ratio
$$x = [1/(\mu~m_{\mathrm H})](kT_{\mathrm s}/V_{\mathrm s}^2)$$
of the mean post-shock temperature to the square of the shock
speed is equal to 3/16. In nonlinear shock acceleration, $x$ drops
well below this value (Ellison 2000, Decourchelle, Ellison, \& Ballet
2000). Thus independent measurements of the shock velocity and
post-shock temperature give a powerful diagnostic of efficient shock
acceleration. The shock velocity can be estimated from X-ray expansion
measurements (see Vink et al. 1998; Koraleski et al. 1998; Hughes
1999; Hughes 2000) and possibly from Doppler shifts of X-ray lines (or
Balmer lines as well), while the post-shock electron temperature can
be derived from spatially resolved X-ray spectra of the downstream
region.  However, as discussed in section 2.2.2 the electron
temperature is most likely not in equilibration with the ion
temperature. The difficulty is then to distinguish the effects of
efficient acceleration from a lack of temperature equilibration
between electrons and ions.

The {\it Chandra} observation of 1E 0102.2-7219 in the Small
Magellanic Cloud has allowed the determination of both the shock
velocity and the post-shock electron temperature as demonstrated by
Hughes, Rakowski, \& Decourchelle (2000).  The well known distance to
the Small Magellanic Cloud allows a better determination of the shock
velocity than for galactic supernova remnants, whose distance is often
not well established.  While the shock velocity is estimated to be
$\simeq 6000$~km/s, the post-shock electron temperature is 0.4-1 keV,
which is almost 25 times lower than expected for a 6000 km/s
test-particle shock.  Hughes et al.  (2000) have shown that even
Coulomb heating alone would produce a higher electron temperature than
observed unless a substantial fraction of the shock energy should have
gone in accelerating particles, although the exact amount remains to
be determined depending on the degree of equilibration between
electron and ion temperatures.

In non-radiative shocks, optical and UV lines yield important
constraints on this degree of equilibration (see section 2.2.2).
Direct X-ray determinations of the ion temperature(s) are in principle
possible, and will be available in future, once instruments have
sufficient spectral resolution to measure the Doppler broadening of
the X-ray lines.

\paragraph{Overall effects on the SNR geometry}

The nonlinear shock modification impacts not only the jump conditions, but
also the overall structure of the remnant. The non negligible fraction of
accelerated ions modifies the compressibility of the shocked gas, and gives
rise to a modified structure of the shocked material (e.g. Chevalier 1983).
In young supernova remnants, the interaction region (between the reverse
shock and the forward shock) gets thinner with higher efficiency acceleration
at the shocks, and has higher densities and lower temperatures as shown by
Decourchelle, Ellison, \& Ballet (2000). The position of the forward and
reverse shocks with respect to the contact discontinuity (interface between
stellar and ambient material) provides information on the efficiency of the
acceleration at each shock (see Decourchelle et al. 2000). X-ray imaging
instruments, which map the whole shocked region (and not simply the post-shock
region), can give information on the location of both shocks and constrain
their respective acceleration efficiency. However, while the forward shock
can be easily observed (see for example 1E 01012.2-7219, Gaetz et al. 2000),
the exact position of the reverse shock is difficult to establish due to
projection effects and density structure in the unshocked ejecta.

\paragraph{Effects from low-energy suprathermal electrons}

Shock acceleration predicts an extended suprathermal tail to the
downstream electron energy distribution instead of the exponential
cutoff characteristic of a thermal Maxwellian (e.g., Bykov and Uvarov
1999; Baring et al. 1999).  The entire electron distribution will
radiate bremsstrahlung photons, an electron with energy $E$ typically
producing a photon of energy $E/3$.  This low-energy end of the
nonthermal electron distribution can also produce potentially
observable effects through collisional ionization and excitation.

The nonthermal-bremsstrahlung continuum should have a concave-upward
curvature.  This has not been seen, either because of the presence of
synchrotron emission, or because thermal bremsstrahlung still
dominates.  In most cases, however, one expects that above about 10
keV any synchrotron component must be dropping away rapidly.  Images
and spectra of SNRs in the range of tens of keV with adequate
sensitivity should certainly find nonthermal bremsstrahlung, whose
analysis will give important direct information on the production of
low-energy cosmic-ray electrons and on the details of the injection
mechanism.

Ionization and excitation rates from electron impact are calculated by
integrating energy-dependent cross sections over the electron energy
distribution and will differ most from those calculated for a strictly
Maxwellian distribution.  Similar calculations have been done for the
solar corona and other stellar coronae (e.g., Owocki \& Scudder, 1983)
and differences in the ionization balance have been found.  Hampering
this effort is our poor understanding of electron injection, so that
there is no unambiguous prediction for the shape of the low-energy end
of the nonthermal electron distribution.  Calculations for power laws
are not unreasonable at this stage. This has recently been done for a
range of possible power laws (Porquet, Arnaud, \& Decourchelle 2001):
the increase of the ionisation rates depends on the fraction of
nonthermal electrons above the ionisation potential and can reach
several order of magnitudes. The ionisation balance can be affected
significantly, in particular at low temperatures, but the effects are
less for the ionizing plasmas expected in young supernova remnants.
The new generation of X-ray spectrometers ({\it Chandra/HETGS} and
{\it XMM-Newton/RGS}) is providing high spectral resolution spectra of
the brightest and smallest angular size supernova remnants like 1E
0102.2-7219 (see Rasmussen et al. 2001), which will allow precise line
diagnostics, relevant for constraining nonthermal ionisation and line
excitation.

\paragraph{Synchrotron continuum}

If no lines are present, or if the continuum steepens and can be shown
not to be thermal by other arguments, synchrotron emission is the most
likely possibility.  Simple considerations (e.g., Reynolds 1996) show
that one can readily expect synchrotron emission to photon energies
above 10 keV from remnants up to and perhaps beyond $10^4$ years in
age.  However, in all SNRs observed in both radio and X-rays, the
X-rays (thermal or not) fall below the extrapolation of the radio
spectrum, indicating that the electron spectrum has begun to steepen
at what turn out to be energies of no more than about 10 TeV in most
cases (Reynolds and Keohane 1999).  While this rolloff may be due to
radiative losses in some cases, for all five historical remnants in
the sample, the radiative-loss limit is higher than the one actually
observed, indicating that the cutoff is due to some other process and
is presumably the same for ions as for electrons.  Unless the older
remnants improve unexpectedly in their ability to accelerate particles
to the highest energies, (perhaps through magnetic field amplification
as recently suggested by Lucek and Bell, 2000) the ability of SNRs to
produce power-law spectra up to the ``knee'' is in question.
As we discuss below, Cas A, because of its extremely high inferred
magnetic field ($\sim 1000\mu$G), should be able to accelerate
protons to well above $10^{15}$ eV (Allen et al. 1997).
 
Where synchrotron X-rays are required to explain part or all of the
observations, power laws have mainly been used for the analysis.  Over
the limited spectral range of current X-ray satellites, this is a
reasonable approximation, though it is not expected physically.  The
sharpest cutoff naturally arising through shock acceleration is an
exponential in electron energy (Webb, Drury, \& Biermann 1984; Drury
1990) and in a real remnant, spatial inhomogeneities and time
dependent effects will result in a rolloff even broader than this.
Extensive models have been calculated in Reynolds (1998).  A few of
the simpler ones are available in the X-ray spectral fitting package
XSPEC (v.11) and have been applied to SN 1006 (Dyer et al.~2000) and
RCW 86 (Borkowski et al. 2001).  Rolloff frequencies found in these
fits imply electron energies of the same order as the upper limits
cited above, of tens of TeV.  Similar results are found in more
complete models (Berezhko, Ksenofontov, \& Petukhov 1999) which
calculate the cosmic ray acceleration self-consistently and match the
broad-band continuum emission including the GeV and TeV gamma-ray
observations (see also Ellison, Berezhko, \& Baring 2000).
 
Morphological information on synchrotron X-rays can also be important.
X-ray emitting electrons have energies of order 100 TeV and, should
diffuse observable distances ahead of the shock as they are
accelerated.  These synchrotron halos are similar to the precursors
expected in radio, but on a larger scale corresponding to larger
diffusion lengths.  Unlike radio halos, they should always be large
enough to resolve, but unfavorable upstream magnetic-field geometry (a
magnetic field primarily along the line of sight to the observer)
might make them unobservably faint.  Behind the shock, X-rays should
come from a narrower region than in the radio because of radiative
losses; however, projection effects may make this effect difficult to
observe.

\subsubsection{Gamma-ray diagnostics}
The radio and X-ray synchrotron emission probes the accelerated
electron population, as do the non-thermal bremsstrahlung and inverse
Compton components.  They tell us nothing directly about the nuclear
component which dominates in the GCR. Detection of gamma-ray emission
from SNRs clearly produced by $\pi^0$ decay would be unambiguous
direct proof of the existence of accelerated nuclei in SNRs (e.g.,
Drury et al, 1994; Naito and Takahara, 1994; Berezhko \& V\"olk 2000).

If TeV gamma rays can be shown (for instance, from spectral evidence)
to be inverse-Compton upscattered CMB photons, the factor $R$ by
which the extrapolation of the radio spectrum to TeV energies
exceeds the TeV flux can give the magnetic field (if synchrotron
and IC occupy the same volume, unlikely to be exactly the case):
$$R = 3.72 \times 10^3 B_{\mu G}^{1.6}.$$
In SN 1006, $R \sim 1.4 \times 10^5$ (Tanimori et al.~1998; Green
1998) implying $\langle B \rangle = 9.6 \ \mu$G.  A more accurate
model involving calculating the electron density everywhere also gives
$\langle B \rangle = 9 \ \mu$G, and implies an efficiency of electron
acceleration of about 5\%.  As more TeV observations of SNRs become
available, similar calculations will be possible for more objects.
Nondetections, or attibutions of TeV emission to other processes such
as $\pi^0$ decay or bremsstrahlung, give lower limits on the mean
magnetic field.

It is important to note that the diffusive shock acceleration
mechanism is expected to place a larger fraction of the available ram
kinetic energy into ions than electrons (perhaps 10 times as much or
more). If electron efficiencies as high as 5\% are inferred for SN1006
and other SNRs, these objects may be producing cosmic rays with very
high efficiencies. If this is the case, test-particle models will be
clearly inadequate and nonlinear models of particle acceleration will
need to be applied to both the SNR dynamics and the particle spectra
(Decourchelle, Ellison, \& Ballet, 2000; Hughes, Rakowski, \&
Decourchelle, 2000)

\subsubsection{Charged particle diagnostics}

There are three aspects of particle spectra that need consideration in
matching cosmic ray observations: (1) the shape, (2) the maximum energy,
and (3) the possibility of features (i.e., bumps) caused by individual
nearby sources dominating the spectrum.

The accelerated particle spectrum at the source is predicted to be at
least as hard or even somewhat harder than an $E^{-2}$ energy spectrum
over a wide energy range.  The actual source spectra inferred from
observations after propagation corrections tend to be softer (steeper)
than this, spectral exponents around 2.1 being favoured by propagation
models with negligible reacceleration and values as soft as 2.4 being
required for the models with significant reacceleration. This is a
worrying discrepancy.

The upper energy limit of acceleration is determined essentially, as
is evident on dimensional grounds, by the product of shock radius,
shock velocity and magnetic field; as the famous Hillas plot shows
this severely restricts the possible Galactic acceleration sites.
Specifically, the maximum energy, $\Emax$, can be estimated by first using
a model of SNR evolution (e.g., Truelove \& McKee 1999 which continuously
maps between the free expansion and Sedov phases) to give the shock
parameters (i.e., speed, $\Vsk$, and radius, $\Rsk$) as a function of
explosion energy, $\EnSN$, ejecta mass, $\Mej$, and remnant age, $\tSNR$.
The maximum momentum where the spectrum cuts off is then estimated by
setting the diffusive shock acceleration time, $\tacc$, equal to $\tSNR$,
or by setting the upstream diffusion length equal to some fraction of the
shock radius, whichever produces the lowest $\Emax$.
In fact, an accurate determination of $\Emax$ in an evolving remnant
requires a more complete model than this (Berezhko, 1996), which keeps
track of the history of particles, adiabatic losses, and the numbers
of particles accelerated at a given epoch. However the results of the
more sophisticated model agree within factors of order 2 with the
simple estimates.

For electrons, $\Emax$ is equal to that of the protons or to the value
determined from combined synchrotron and inverse-Compton
losses, whichever is less (see Baring et al. 1999 for details).

Recent studies of SN1006 (Reynolds 1996; Berezhko et al. 1999;
Ellison, Berezhko, \& Baring 2000) and Kepler's SNR (Ellison 2000)
indicate that, while these SNRs accelerate particles to TeV energies, they
do not produce particle energies anywhere close to $10^{15}$ eV.  Cas
A on the other hand, because of its extremely large magnetic field
($B\sim 1000 \mu$G) should be easily able to produce particles with
energies of $10^{15}\,\rm eV$ or higher.  This suggests that only some
subclass of SNRs can provide the knee particles while most SNRs have
spectra cutting off at considerably lower energies (Reynolds and Keohane
1999). This, in turn, suggests that features may be observable in the
GCR spectrum even well below the knee and that the number of local
sources contributing to the knee region may be quite small (cf Erlykin
and Wolfendale, 2000).

\section{CR Compositional tests}

There has been much debate about the origin of the material
accelerated to form the GCR. The inferred chemical composition of the
GCR at source (that is after slightly model-dependent corrections for
propagation) is now rather well determined for all the major species
and even many minor ones.  The composition is basically quite close to
solar, but with significant differences as reviewed in J-P Meyer's
talk.

\subsection{Acceleration fractionation effects}

The nonlinear shock acceleration models make quite specific
predictions about the composition of the accelerated particles
compared to the composition of the medium into which the shock is
propagating.  Firstly, any seed population of pre-existing nonthermal
particles in the upstream medium will be picked up and accelerated by
the shock with essentially no fractionation (this is basically the
original linear test-particle picture of shock acceleration).
Secondly, and much more interestingly, the nonlinear theory requires
that the shock-heated particle distribution contain a nonthermal tail
extending to very high energies. The transition from the thermal
population to the nonthermal tail defines what is usually called the
``injection'' rate.  This is perhaps the most important point about
nonlinear shock acceleration, {\em there is no need for a separate
injection process}, a shock propagating in a given medium accelerates
particles out of that same medium. In fact the important point about
this injection process is not that it is difficult, but rather, as has
been emphasised recently by Malkov, that it is too easy. The problem is
that there is simply not enough energy to accelerate very many particles
to relativistic energies. Thus the nonlinear reaction effects on the
shock structure and the dissipative processes operating in the
subshock have to conspire to throttle back the effective injection
rate to a reasonable value. In the case of SNR shocks this means that
the effective injection rate for protons has to be of order
$10^{-4}$.

The protons are the key species because they dominate the energy
budget. Because the shock is a collisionless plasma shock dominated by
magnetic field effects the throttling back must be done by what is
basically a gyroradius filtering effect whereby the injection of
particles with magnetic rigidity of order that of a shock-heated
proton is strongly suppressed. However particles of higher rigidity
and larger gyroradius do not interact as strongly with the small scale
fields and structures responsible for this suppression and therefore
are more efficiently injected. Although the details are complicated,
and not really understood, there is a clear qualitative
prediction. For a shock propagating in a multispecies medium, but one
dominated in mass density by hydrogen, the compostion of the
accelerated particles relative to the preshock medium should show a
fractionation which is a smooth monotonically increasing and then
saturating function of initial particle mass to charge ratio (this is
often referred to as an $A/Q$ effect
and, in fact, the $A/Q$ enhancement may not be strickly monotonic for
extremely low Mach number shocks, e.g., Ellison, Drury, \& Meyer
(1997)).
Of course this refers only to the nuclear and other heavy species
with mass to charge ratios greater than that of the proton; the problem of
the electron injection rate is much more complicated.

Qualitatively, this fits the general overabundance of heavy elements
in the GCRS composition relative to solar, but it is impossible to get
a good quantitative fit using equilibrium ionisation models of a gas
phase ISM with standard composition
(however if only volatile elements are considered, $A/Q$
does allow a good fit; \cite{Ellison97}, \cite{Meyer97})

\subsubsection{Dust}

In much of the ISM the refractory elements are not in the gas-phase
but condensed into small interstellar dust grains. These grains are
electrically charged and will behave like very heavy ions. Because of
their enormous gyroradii they are injected with essentially unit
efficiency, but are only very slowly accelerated. Detailed estimates
of the acceleration and other length and time scales show that the
accelerated dust particle distribution will cut-off at dust velocities
about ten times that of the shock because of frictional losses. The
collisions between the gas atoms and the dust grains which are
responsible for this friction also induce a certain amount of
sputtering of secondary ions from the grain surface, and because the
grains are diffusing on both sides of the shock some of these
secondary ions are produced in the upstream region just ahead of the
shock. Detailed estimates show that, independent of dust, gas and
shock parameters, this yields a seed ion population upstream which is
$O(10^{-3})$ suppressed relative to the top end of the accelerated
dust distribution. These ions are swept into the shock and accelerated
with little or no further fractionation, and because the protons are
suppressed relative to the bulk material by about $10^4$ the final
effect is that the sputtered dust component is expected to show
roughly an order of magnitude enhancement relative to hydrogen with
little fractionation.

\subsubsection{Ionization state}

The volatile species accelerated out of the gas phase, on the
contrary, are expected to show a strong $A/Q$ fractionation
effect. The problem here is to estimate the effective charge state of
these species. Except in the very hot phases of the ISM it is unlikely
that the mean charge is more than $+1$ or $+2$, however in the shock
precursor region photo-ionization by radiation from behind the shock
is probably also important. This is one area where a detailed study
would be very useful. For the moment if one simply assumes that the
volatile ISM species are unlikely to have lost more than one or two
electrons, the mass can be taken as a proxy for the effective $A/Q$
value. The prediction of shock acceleration out of a dusty ISM is then
that two components should be visible in the compositional data. The
volatiles should lie on a rather smooth fractionation line which is a
monotonically increasing function of atomic mass. In addition, there
should be a refractory component from dust which shows little or no
mass dependent fractionation, but which is globally enhanced by a
factor of about ten relative to hydrogen.  This appears to be in very
good agreement with {\em all} the data on the chemical composition of
the GCRS. In particular, 10\% of oxygen must be in grains
from dust chemistry, and we know that substantial amounts of
carbon are also in grains as well as the gas phase so that these two
elements should fall between the two bands, exactly as observed.

\subsubsection{FIP}

The alternative ``explanation'' of the pattern of elemental abundance
variations observed between the GCRS and solar is the so-called FIP
effect. It is known that in the solar corona elements with first
ionization potentials (FIP) below about $10\,\rm eV$ are enhanced by
about an order of magnitude and there is evidence that the same effect
occurs in the coronae of other cool stars. This effect biases the
composition of solar energetic particles and produces a composition
which is remarkably similar to that of the GCRS. This has lead to
suggestions that the source of the GCR material is to be found in
coronal mass ejections from solar-like stars, but it is difficult to
make a plausible quantitative model along these lines. Somehow one has
to produce a large sea of low energy particles from dwarf stars which
survive long enough to encounter strong SNR shocks and be accelerated
without being swamped by fresh particles accelerated by the shock out
of the ISM. The close resemblence of a FIP-biased composition to that
predicted for particles accelerated by a shock from a dusty ISM
results of course from the fact that FIP correlates strongly with
chemistry. The reactive elements which form strongly bound
chemical compounds and condense out of the gas phase are mostly those
with low first ionization potentials.

This is often presented as a FIP versus volatility debate, but this is
not really correct. The comparison should be between the specific
physical model of acceleration by an SNR shock of particles from a
dusty ISM with standard bulk composition and grain chemistry, and a
more speculative scenario of injection of FIP-biased material from
young dwarf stars with high levels of flaring activity followed by
subsequent SNR shock acceleration of this material.

\subsection {$^{22}$Ne and Wolf--Rayet stars}

There is now strong observational evidence that the (isotopic)
composition of the Galactic Cosmic Rays (GCRs) exhibits some
significant deviations with respect to the abundances measured in the
local (solar neighbourhood) interstellar medium (ISM) (see Lukasiak et
al.  1994; DuVernois et al.  1996; Connell \& Simpson 1997; George et
al.  2000; Wiedenbeck 2000).  The most striking difference between the
isotopic composition of the Galactic Cosmic Ray Sources (GCRS) and the
solar system is the factor $\sim$ 3 enhanced $^{22}$Ne/$^{20}$Ne ratio
observed in the GCRS, while isotopic ratios involving heavier isotopes
like magnesium or silicon have near solar values.  Wolf--Rayet (WR)
stars appear as a promising source for the $^{22}$Ne excesses. 
Indeed, during the WC phase, a particular stage during the WR phase
(see e.g. the review on the WR stars by Willis 1999), stellar models
predict that $^{22}$Ne is significantly enhanced in the stellar winds. 
Let us emphasize here that the predicted $^{22}$Ne excesses in the
winds of the WC stars have been recently confirmed observationally
(Willis et al.  1998; Dessart et al.  2000).

Two scenarios have been proposed in order to account for the
differences (`anomalies') in the isotopic composition of the GCRSs. 
In both, WR stars play an important role as the source of the
$^{22}$Ne excess.  The first scenario invokes two distinct components
to be accelerated to GCR energies (e.g. Arnould 1984; Prantzos et al. 
1987, and references therein).  The first component, referred to as
the `normal component', is just made of ISM material of `normal' solar
composition, while the other one emerges from the nuclear processed
wind of massive mass--losing stars of the WR type, and is referred to
as the `wind component'.  It has been demonstrated that this type of
scenario is able to account in a natural way for the excess of
$^{22}$Ne.  Typically, the fractional contribution of the wind
component to the bulk GCRs found by imposing the model reproduction of
the observed GCR $^{22}$Ne/$^{20}$Ne ratio (adopted here equal to 3)
ranges between about 2 and 10\%, depending upon the model star (see
the contribution by Meynet et al.  in this volume).

A second scenario has been proposed in order to explain the GCR
composition anomalies (Woosley \& Weaver 1981; Maeder 1984; Meynet \&
Maeder 1997; Meynet et al.  in this volume).  This model, envisions
the acceleration of ISM material whose composition is different from
the normal one used for comparison (i.e. the ISM in the solar
neighbourhood).  More specifically, it is assumed that the accelerated
ISM originates from the inner regions of the Galaxy, where the star
formation and supernova rates are higher than in the solar
neighbourhood.  As a consequence, the metallicity is higher and the
ISM isotopic composition is very likely to be different as well.

Whilst several observational data can be accounted for, both models
still face difficulties.  In the two--components scenario, it remains
to be seen if the WR wind component can be accelerated with a large
enough efficiency in order to contaminate at a high enough level the
normal component made of ISM matter of typical local (solar
neighbourhood) composition.  The metallicity--gradient model faces
more specific problems related to the construction of reliable
chemical evolution models of the Galaxy, and in particular to the
predictions of composition gradients in the galactic disc.

On the observational side, further data, concerning in particular
heavy s--process nuclides, would certainly be very helpful in
constraining the models.

\section{Superbubbles as GCR sources}

As recalled above, energetic considerations make SN explosions a very
probable energy source for GCRs.  SN explosions, however, are not
random in the Galaxy, and show strong spatial and temporal
correlations resulting from the concentration of the vast majority of
(core-collapse) SN progenitors into OB associations, formed on a short
timescale from the collapse of a giant molecular cloud (e.g. Blitz,
1993).  The explosion of the first SN among such an association is
thus to be followed by several tens of others within a few million
years, at approximately the same location.  This results in the
formation of a `multiple supernova remnant', powered by both the SN
explosions and the strong winds of Wolf-Rayet stars in the OB
association, which grows as a large bubble of hot, tenuous plasma
known as a superbubble (e.g. MacLow \& McCray, 1988; Tomisaka 1998;
Korpi et al.  1999).  Superbubbles (SBs) are commonly observed in
X-rays in our and nearby galaxies.

The impact of multiple SNe on their environment is large, and if SN
explosions are indeed the GCR source of energy, the fact that most SNe
occur in groups should incite us to take their collective effect into
account when considering cosmic ray acceleration.  In particular, it
seems natural to expect an intense supersonic turbulence inside the
accelerator, due to the interaction of individual SN shocks and strong
stellar winds in the SBs.

\subsection{Particle acceleration inside superbubbles}

The theory of diffusive shock acceleration described above applies to
the `regular' shocks of distinct, isolated SNRs.  For most SN
explosions occuring inside SBs, however, the `well-ordered SNR' stage
may be relatively short due to the interaction with the pre-existing
strong turbulence (primary and secondary shocks from previous SNe and
stellar winds).  Assuming a length scale of about 10~pc between two
major shocks, and a large ambient magnetic field strength $\gtrsim 30$
$\mu$G, the timescale for the disruption of a SN shock inside a SB can
be roughly estimated as $\sim 1000$ years.  Given the very low ambient
density ($\lsim 10^{-2}\,\mathrm{part.cm}^{-3}$), only a small amount
of matter can flow through the forward shock of a SN before it
encounters another strong shock or a clump of denser material and
generates a series of secondary shocks by reflection, eventually
leading to the mentioned strong turbulence.  The mass contained inside
a sphere of 10~pc with a density of $10^{-2}\,\mathrm{cm}^{-3}$ is
only 1 solar mass!  However, particle acceleration does not cease when
the strong magnetic turbulence develops.  On the contrary, it is
expected to become very efficient, and SBs have been considered as
very plausible sites of nonthermal particle acceleration (Bykov \&
Fleishman, 1992; Parizot, 1998; Higdon, Lingenfelter \& Ramaty, 1998).

The acceleration mechanism in SBs should enable an efficient
conversion of the MHD energy of large scale shocks and plasma motions
into CRs.  The SB acceleration model is based on the kinetic equation
for the particle distribution functions in the stochastic velocity
field with multiple shocks inside SBs.  It has been developed by Bykov
\& Toptygin, 1987; Bykov \& Fleishman, 1992; Bykov, 1999.  The models
account for the creation of a nonthermal population of nuclei with a
hard low-energy spectrum, containing a substantial part of the kinetic
energy released by SNe and massive stellar winds.  Test particle
calculations pointed at the importance of nonlinear effects, and a
nonlinear model accounting for the reaction of the accelerated
particles on the shock turbulence inside the SB has thus also been
developed by Bykov (1999) (see also Bykov 2000, this volume), the
outcome of which is a strong temporal evolution of the energetic
particle spectrum.  The energy contained in the superbubble MHD
turbulence is converted into nonthermal nuclei with an efficiency
estimated as $\gsim$ 20 $\%$.

\subsection{Superbubbles and light element production}

Up to now, the SB acceleration model has been mostly applied to light
element nucleosynthesis and Galactic evolution.  Among the light
elements (Li, Be and B), the isotopes $^{6}$Li, $^{9}$Be and $^{10}$B
(and maybe $^{11}$B as well) are produced exclusively by spallation of
heavier nuclei, mostly C and O. Recent studies have shown that the
original Galactic nucleosynthesis scenario (Reeves et al., 1970;
Meneguzzi et al., 1971) in which CR protons interact with C and O
nuclei in the ISM was much too inefficient in the early Galaxy (where
C and O are very rare).  On the other hand, the possibility of
accelerating particles out of the enriched material inside SBs, filled
with stellar wind and SN ejecta, led people to consider the so-called
\textit{superbubble model} for Li, Be and B production (see the
articles by Parizot and Ramaty in the present volume, and references
therein).  This model proved capable of accounting for all the current
observational constraints pertaining to light element nucleosynthesis
and evolution, which no other known mechanism can do.  In particular,
particle acceleration at the shocks of individual SNe, even when
taking into account the ejecta accelerated at the reverse shock, is
unable to explain the observed very efficient production of Li, Be and
B in the early Galaxy (Parizot \& Drury, 1999a,b).

Now if the collective effects of multiple SNe inside SBs dominate in
the early Galaxy, it is natural to ask whether this is not also the
case in present times, since most SNe do explode inside SBs.  This
would result in a harder energy spectrum at low energy (which
incidentally makes the spallation reactions more efficient), with most
of the system energy being imparted to particles in the 0.1--1~GeV/n
range.  Another consequence would be a substantial enrichment of the
cosmic rays by freshly synthesized nuclei, from SN ejecta and
Wolf-Rayet stellar winds.  This would offer a natural way to solve the
Ne isotopic ratio problem.

\subsection{Superbubbles and GCRs}

Most of what has been acquired from the study of CR acceleration by
diffusive shock acceleration in isolated SNRs applies also in the
context of superbubbles.  According to the model developed by MacLow
\& McCray (1988) for the dynamical evolution of a SB, the material
inside a SB (and thus entering the acceleration process) is composed
mostly of the swept-up ISM, contaminated by a few percent of enriched
material of stellar origin.  This is in very good agreement with the
conclusions of the study of light element nucleosynthesis (Parizot,
2000, and this volume).  Therefore, according to this scenario, the
basic material from which the CRs are accelerated is close to the mean
ISM, just as in the scenario involving individual SNRs (although the
composition inside SBs was much richer than the mean ISM in the early
Galaxy).  Most importantly, the SB lifetime and density are too small
for the ionisation equilibrium to be reached and the dust grains are
probably not destroyed in the imparted time, in spite of the very high
ambient temperature.  Therefore, the mechanism described above (and in
Ellison et al., 1997) resulting in the enhanced injection of
refractory elements into the acceleration process should work equally
well in the ISM and at the strong shocks inside SBs.  However, if most
of the GCRs originate from SBs, then the `abnormally' high abundance
of $^{22}$Ne among GCRs may be more easily understood.  Indeed, Meynet
has shown that the Neon isotopic ratio observed among GCRs requires
that about $6\%$ of the accelerated material is made of Wolf-Rayet
winds (see Meynet's article, this volume).  This is remarkably close
to what is expected inside SBs, from the study of both SB dynamics and
light element nucleosynthesis.

In addition, superbubbles might provide an acceleration mechanism
drawing its energy from the SN explosions, just like the standard SNR
model, but allowing for a higher energy cut-off, extending beyond the
knee, notably because of their larger dimension.  The problem of VHE
CR acceleration in the superbubble model, has been addressed by Bykov
\& Toptygin (1997) in the framework of CR acceleration by multiple
shocks.  They estimated the maximal energies of accelerated nuclei as
$\sim 10^{18}$~eV, in the presence of amplified fluctuating magnetic
fields in the bubble interior of order 30~$\mu$G. In this model, the
spectrum beyond the knee and up to 10$^{18}$~eV is dominated by heavy
nuclei.

\subsection{Observational evidence}

The SB acceleration mechanism appears natural from the theoretical
point of view, since most of the SNe are indeed known to occur in this
kind of environments, and is supported by the study of LiBeB
production and evolution in the Galaxy.  Interestingly enough, it now
seems to be receiving direct observational support as well.

Recent observations of interstellar Li abundances (Knauth et al.,
2000) have shown evidence of newly synthetized lithium in the Perseus
OB2 Cloud, the value of the $^{6}$Li/$^{7}$Li ratio being found both
10 times higher than in the standard ISM and very close to the
spallation ratio.  This could indicate that a very significant Li
production by spallation has recently taken place about the Perseus
OB2 association, on a timescale shorter than the chemical
homogenization timescale.  The huge Li production required to locally
overtake the standard Li production accumulated since the beginning of
Galactic chemical enrichment could be explained within the SB model,
and pleads in favour of an efficient particle acceleration inside
superbubbles.

Another observation by Cunha et al.  (2000) reported an unexpected
anti-correlation of B and O abundances in the Orion OB1 association.
Since both $^{16}$O and $^{11}$B are thought to be produced by SNe,
one would expect instead a positive correlation of their abundances.
However, Cunha et al.  (2000) have shown that the anticorrelation can
be explained if a strong B production by spallation has recently
occured in the neighbourhood.  Since the Orion clouds are just at the
border of a typical example of a SB being blown by repeated SNe, the
unexpected B-O anticorrelation would be rather natural in the context
of the SB model for particle acceleration.

A confirmation of this interpretation of the B-O anticorrelation in
Orion is in progress, through the observation of the Be abundance in
the same stars.  Conclusive evidence for the particle acceleration
inside SNs would be provided by the observation of X-ray and gamma-ray
line emission from the SB shell and/or neighbouring molecular clouds.
The detection of these K lines and nuclear de-excitation lines,
possibly with INTEGRAL, would also allow one to determine more
precisely the composition and spectrum of the SB energetic particles,
and provide strong constraints on the SB acceleration mechanism.

\section{Nasty Problems}

The standard picture, that the bulk of the GCR originate from shock
acceleration associated with strong SNR shocks, can justifiably claim
a number of notable successes, however there remain a number of
nasty problems which we wish, in conclusion, to point out ({\it cf}
also Kirk and Dendy, 2001).

\subsection{O-star wind termination shocks}

A decade ago Lozynskaya (1991) pointed out that the terminal shocks in
the winds from O-stars are very similar to SNR shocks, yet there is no
evidence for non-thermal effects associated with O-star wind
bubbles. What causes this difference?

\subsection{The knee and beyond}

The standard picture makes a clear prediction that the GCR spectrum
should start to cut-off at rigidities of about $10^{14}\,\rm V$ or
less for all species and drop exponentially as one goes to higher
energies.  The data, on the contrary, shows only a very slight
feature, the famous ``knee'' starting at about $10^{15}\,\rm V$ and
continues to at least the ``ankle'' region of $10^{18}\,\rm V$.

\subsection{Soft source spectra}

The nonlinear acceleration models do not produce precise power-law
spectra, but they do put roughly as much, if not more, energy per
logarithmic interval into the region near the upper cut-off as into
the region around $1\,\rm GV$ where the protons are mildly
relativistic.  Thus the effective differential energy spectral index
is close to 2.0, distinctly harder than the 2.3 to 2.4 range favoured
by reacceleration models for Galactic propagation.

\section{Conclusions}

The prospects for interesting science are very good. On the one hand,
observing capabilities are improving rapidly. On the other, the models
are making definite predictions of potentially observable effects. And
as the list of ``nasty'' problems shows, there is much that we do not
understand. Contrary to the general ``folklore'' it is by no means
certain that SNRs are the source of the GCR and in fact the existence
of the ``knee'' and the particles above the ``knee'' is fairly clear
proof that something else is required.

\end{article}
\end{document}